\documentclass{aip-cp}

\usepackage[numbers]{natbib}
\usepackage{rotating}
\usepackage{graphicx}
\usepackage{subcaption}
\usepackage{breqn}
\usepackage{amsfonts}
\usepackage[section]{placeins}
\usepackage[T1]{fontenc} 

\begin{document}

\title{Test of an Analytic, Finite, Non-Perturbative, Gauge-Invariant QCD formulation against elastic pp scattering at the ISR energies.}

\author[aff1]{Peter H. Tsang}
\eaddress[url]{http://www.aip.org}
\eaddress{peter\_tsang@brown.edu}

\affil[aff1]{Physics Department, Brown University, Providence, Rhode Island, USA.}

\maketitle

\begin{abstract}
A recent QCD formulation that is non-perturbative, finite, gauge-invariant, exact emerged from Schwinger's Generating Functional. A first test of the validity of this formulation is provided here against elastic proton-proton scattering at the Intersecting Storage Rings (ISR) energies. Extension to LHC energies is currently underway.
\end{abstract}

\section{INTRODUCTION}
 A recent formulation by H.M. Fried$^1$, Y. Gabellini, T. Grandou\footnote{presenting at this conference.}, R. Hofmann$^1$, Y.M. Sheu, P.H. Tsang$^1$, obtained analytic, finite, gauge-invariant, non-perturbative equations for QCD processes, \cite{qcd1}-\cite{qcd6}.
Starting with the Schwinger Generating Functional, 
\begin{equation}
  \mathcal{Z}_{QCD}[j,\bar{\eta},\eta]=\mathcal{N}e^{-\frac{1}{2}\int \frac{\delta}{\delta A}\cdot D_c^{(0)}\cdot \frac{\delta}{\delta A}} \cdot e^{-\frac{i}{4}\int \textbf{F}^2 + \frac{i}{2}\int A\cdot (-\partial^2)\cdot A}\cdot e^{i\int \bar{\eta}\cdot \textbf{G}_c[A]\cdot \eta+\textbf{L}[A]} |_{A=\int \textbf{D}_c^{(0)}\cdot j}
  \label{qcdgf}
\end{equation}
where the quark line, $\textbf{G}_c[x,y|\textbf{A}]= [m+\gamma\cdot (\delta-igA\tau)]^{-1}$
and the virtual quark loop, $\textbf{L}[\textbf{A}]=\ln [1-i\gamma \textbf{A}\tau_c[0]]$,
using a reprocity relation \cite{qcd1}-\cite{qcd6}, the $\textbf{F}^2$ field strength is then replaced with Halpern's linear expression in $\textbf{F}$, $e^{-\frac{i}{4}\int \textbf{F}^2} = N \int d[\chi]e^{\frac{i}{4}\int \chi^2 + \frac{i}{2}\int F\cdot \chi}$ \cite{halpern1}. 
Fradkin's gaussian representation for $\textbf{G}[\textbf{A}]$ and $\textbf{L}[\textbf{A}]$  is used \cite{fradkin1}\cite{fradkin2}. 
Experimentally, quarks do not have static transverse coordinates, and thus, the quark coordinates are to be written as a distribution, $\int d^4 x \ \bar{\psi}(x)\gamma_{\mu}A_{\mu}^a(x) \tau_a \psi(x)$, 
where $
  \int d^2 x'_{\perp}\int d^4 x\ \underline{a(x_{\perp}-x'_{\perp})} \ \bar{\psi}(x')\gamma_{\mu} A_{\mu}^a(x) \tau_a \psi(x')$,
with $a(x_{\perp}-x'_{\perp})$ real and symmetric, and $x'_{\mu} = (x'_{\perp},x_L,x_0)$.
The probability of finding two quarks separated by a transverse (or impact parameter) distance is then $f(b)=\int \frac{d^2q}{(2\pi)^2}\ e^{iqb}\ |\tilde{a}(q)|^2$, where we choose $f(b)=f(0)\ e^{-(\mu b)^{2+\xi}}$ with deformation parameter $\xi$ real and small\footnote{H.M.Fried's talk in this conference will go into details.}.

All QCD processes can now be expressed as gaussian differential operations acting upon gaussian potentials in $A$, \cite{qcd1}-\cite{qcd6},
\begin{dmath}
    e^{\mathfrak{D}_A}F_1[A]F_2[A] = \mathcal{N}\ exp[-\frac{i}{2}\int \bar{Q}\cdot (\hat{-(f\cdot \chi)}^{-1}) \cdot \bar{Q}+\frac{1}{2} \mathrm{Tr} \ln (\hat{-(f\cdot \chi)})]\\
    \quad \cdot exp[\frac{i}{2}\int \frac{\delta}{\delta A}\cdot (\hat{-(f\cdot \chi)}^{-1}) \cdot \frac{\delta}{\delta A} - \int \bar{Q}\cdot (\hat{-(f\cdot \chi)}^{-1}) \cdot \frac{\delta}{\delta A}]  \nonumber \\
    \quad \cdot exp[L[A]]
    \nonumber
\end{dmath}
where $\mathfrak{D}_A = exp [-\frac{i}{2} \int \frac{\partial}{\partial A} D_c^{0} \frac{\partial}{\partial A} ]$.
As can be seen, QCD processes calculated from this formulation is gauge-invariant, non-perturbative, finite and exact. A new important property of Effective Locality\footnote{T. Grandou's talk in this conference will explain Effective Locality in details.} emerged and will be discussed in detail by T. Grandou. All gluons exchanged, called the Gluon Bundle ($\it{GB}$), between two quarks are summed as $(f\cdot \chi)^{-1}$. $G[A]$, $L[A]$ and $(f\cdot \chi)^{-1}$, become the elements out of which all QCD processes can be calculated.

\section{Renormalization and sum of all physical processes}
\begin{figure}[!htb]
  \centerline{\includegraphics[width=15cm]{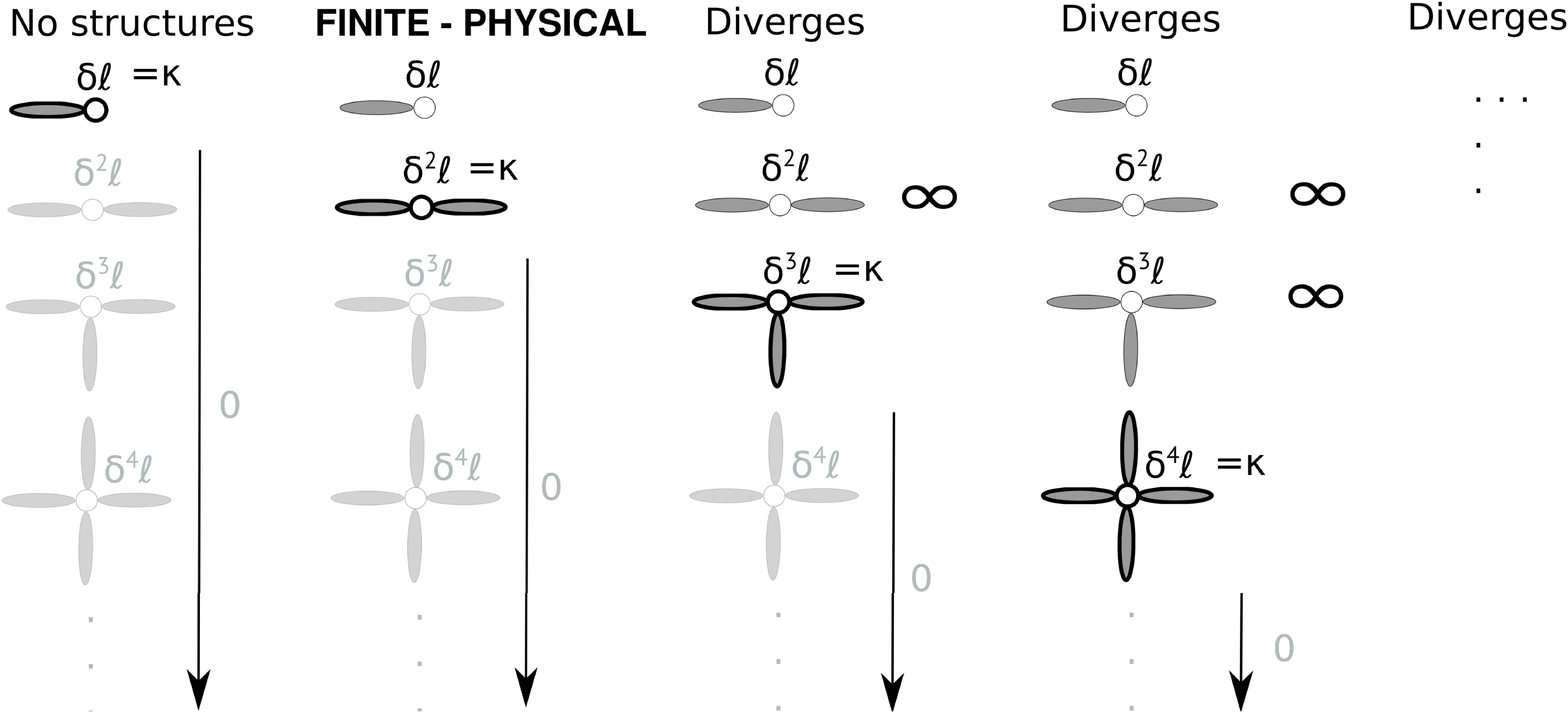}}
  \caption{The only choice for $n$ in $\delta^n \cdot \ell$ = finite that results in physical structures is $n=2$. This case is the second column, where two Gluon Bundles ($\it{GBs}$) and connected to a single closed-quark-loop ($\it{CQLs}$).}
  \label{renormalization}
\end{figure}
In order to make comparisons with experiments, Gluon Bundles ($\it{GB}$) and closed-quark-loops ($\it{CQLs}$) need to be renormalized. The $\delta$ inside the Fradkin representation where a gluon bundle connects to a closed-quark-loop are to vanish while each closed-quark-loop is UV log divergent, or $\ell$ and $\rightarrow \infty$. Renormalization becomes a problem of finding a value for $n$ such that $\delta^n \cdot \ell$ becomes finite. As shown in Figure~\ref{renormalization}, in the first column where $n=1$, all subsequent structures composed using $n=1$ results in graphs identical to $0$. 
\begin{figure}[!htb]
  \centerline{\includegraphics[width=15cm]{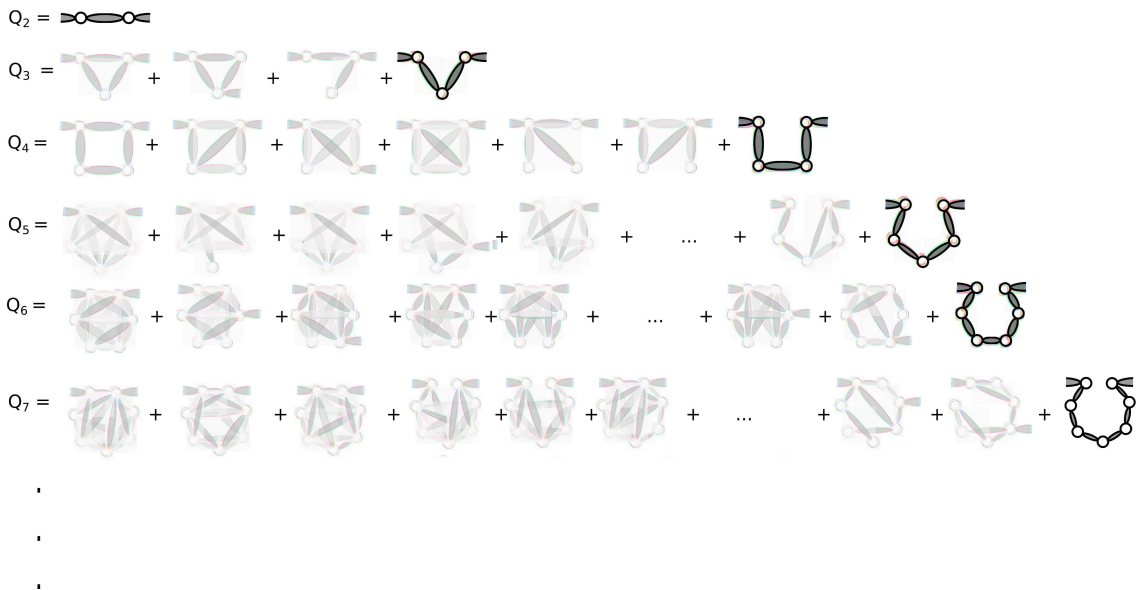}}
  \caption{Renormalization with $n=2$ where $\delta^n \cdot \ell \equiv \kappa$ = finite. Only linear chain-loop-terms become physical and non-zero. All other graphs are explicitly zero, $0$.}
  \label{cluster}
\end{figure}
 In the third column where $n=3$, graphs with fewer  than 3 vertices are divergent while graphs with more than 3 vertices are exactly $0$. All graphs at the 4th column or beyond will only result in divergent quantities.
Only at $n=2$, the second column, results in physical amplitudes. It is with this choice $n=2$ for $\delta^n \cdot \ell = \kappa$ finite that we proceed to compare with experiments. 

The cluster expansion in Figure~\ref{cluster} is used to express the summation of all configurations of $\it{GBs}$ and $\it{CQL}$, \cite{qcd4}. With our renormalization scheme of $\delta^2 \cdot \ell = \kappa$ where $\kappa$ is finite, we can see that for each term of the cluster expansion $Q_i$, only the chain-loop-terms (in bold) are non-zero. Thus, the entire amplitude becomes a finite geometric sum of all the chain-loop-terms as shown in Figure~\ref{cluster}.

At this stage, a two-body approximation instead of the full six-quark problem is used for ease of computation. In the two-body approximation, a phenomenological energy dependence of  $(m_{ext}/E)^{2p}$, is used, with the understanding that it can be derived in the full six-body case. The resulting differential cross-section from summing all the $\it{GBs}$ and chain-loop-terms in Figure~\ref{cluster} is then,
\begin{equation}
  \frac{d\sigma}{dt}(E,q^2) = K\bigg[g^2\beta\ \Big(\frac{m_{ext}}{E}\Big)^{2p}\bigg]^2\bigg[ \frac{1}{4\pi}(9\times 3\times 4)\ \Big(\frac{m_{ext}}{E}\Big)^{2p}e^{\displaystyle{-(3q^2/8m_{ext}^2})} \,-\,\frac{(9\times 3\times 8)\ A_{ext}(q^2)}{1+\beta^2g^2A^2_{int}(q^2)}\bigg]^2
 \label{eq:1}      
 \end{equation}
where $A_{int/ext}(q^2)=\kappa \,(q^2/m_{int/ext}^2)\,e^{\displaystyle{-(q^2/4m_{int/ext}^2)}}$. The coefficients $(9\times 3 \times 4)$ and $(9
\times 3\times 8)$ are for all possible ways two Gluon Bundles, and chain-loop-terms, respectively, can connect between the 3 quarks from one proton to the 3 quarks from the second proton, while preserving color charge. $\beta$ is the sum over the $SU(3)$ angles, $g$ is the coupling constant, $K$ is the conversion factor from $GeV^2$ to $millibarns$.
\section{Comparing theory with ISR experiments}
\begin{figure}[!htb]
    \centering
        \includegraphics[width=\textwidth]{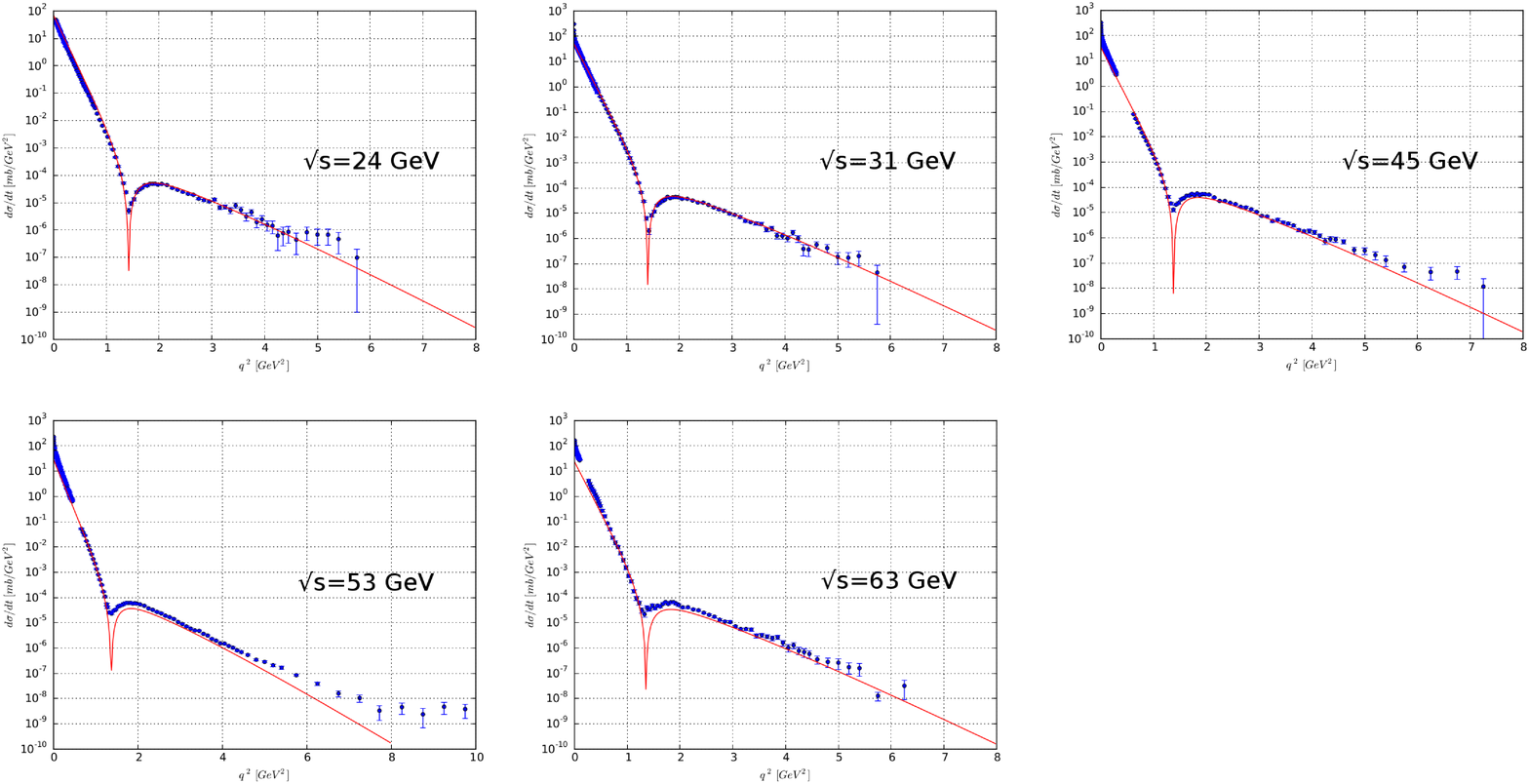}
        \label{isrfits}
    \caption{Differential cross section of elastic pp scattering at ISR energies from 24 GeV to 63 GeV}
\end{figure}
The first term on the R.H.S of Equation~\ref{eq:1} results from Gluon Bundles exchanged between the quark of one proton with the quark of the other proton. This contributes to the left part (left of the diffraction dip) of the ISR curves, Figure~\ref{isrfits}.
The second term on the R.H.S. of Equation~\ref{eq:1} results from the summation of all chain-loop-terms. 

We list the values of the fixed parameters of Equation~\ref{eq:1}:
$K=0.44$ mb GeV$^{-2}$, $g=7.6$, $\beta = 0.30$, $m_{ext} = 0.28$ GeV $\simeq 2m_{\pi}$, $m_{int} = 0.44$ GeV $\simeq 3m_{\pi}$, $p = 0.14$ and $\kappa = 5.22 \,10^{-6}$

We expect that, with the exception of $K$ and $\kappa$, these parameters may have a slight dependence on energy, as it increases from ISR to LHC values, and higher; and that such changes would be due to our two-body approximation of this six-quark scattering reaction.


\section{ACKNOWLEDGMENTS}
This work is supported by a grant from The Julian Schwinger Foundation.



\end{document}